\PassOptionsToPackage{hyperfootnotes=false}{hyperref}
\documentclass[fleqn,usenatbib]{mnras}
\usepackage{newtxtext,newtxmath}
\usepackage[T1]{fontenc}
\usepackage{graphicx,amsmath,xspace,xcolor,needspace,lastpage}
\DeclareRobustCommand{\VAN}[3]{#2}
\let\VANthebibliography\thebibliography
\def\thebibliography{\DeclareRobustCommand{\VAN}[3]{##3}\VANthebibliography}
\newcommand{\br}[1]{\left({#1}\right)}
\renewcommand{\sq}[1]{\left[{#1}\right]}
\newcommand{\kpc}{{\!\;}\rm{kpc}}
\newcommand{\Mpc}{{\!\;}\rm{Mpc}}
\newcommand{\msun}{{\!\;}\rm{M}_{\odot}}
\newcommand{\lsun}{{\!\;}\rm{L}_{\odot}}
\newcommand{\mlsun}{{\!\;}\rm{M}_{\odot}/\rm{L}_{\odot}}
\newcommand{\ud}[1]{{\rm d}{#1}}
\newcommand{\pc}{{\!\;}\rm{pc}}
\newcommand{\mug}{{\!\;}\mu\rm{G}}
\newcommand{\km}{{\!\;}\rm{km}}
\renewcommand{\sec}{{\!\;}\rm{sec}}
\renewcommand{\eqref}[1]{Eq.~\ref{#1}\xspace}
\newcommand{\figref}[1]{Fig.~\ref{#1}\xspace}
\newcommand{\secref}[1]{Sec.~\ref{#1}\xspace}
\newcommand{\hihe}{{H{\sc i}+He}\xspace}
\newcommand{\hi}{{H{\sc i}}\xspace}
\newcommand{\ngc}{{NGC\!\;5055}\xspace}
\definecolor{revisionblue}{RGB}{0,70,165}
\definecolor{literaturepurple}{RGB}{0,0,0}%\definecolor{literaturepurple}{RGB}{110,0,140}
\newif\ifshowchanges
\showchangestrue %\showchangesfalse

\DeclareRobustCommand{\newcitep}[1]{%
  {\ifshowchanges\color{literaturepurple}\fi\citep{#1}}}
\DeclareRobustCommand{\newcitet}[1]{%
  {\ifshowchanges\color{literaturepurple}\fi\citet{#1}}}
\newcommand{\newreference}[1]{%
  \expandafter\def\csname newref@#1\endcsname{1}}
\newreference{Jalocha2012a}
\newreference{Jalocha2012b}
\newreference{Sanchez2004}
\newreference{Sanchez2013}
\newreference{Elstner2014}
\newreference{Knapik2000}
\newreference{Patra2020}
\newreference{Khademi2023}
\newreference{Joung2006}
\let\originalbibitem\bibitem
\renewcommand{\bibitem}[2][]{%
  \color{black}\originalbibitem[#1]{#2}%
  \ifshowchanges\ifcsname newref@#2\endcsname
    \color{literaturepurple}\fi\fi}
\hypersetup{hidelinks,hypertexnames=false}
\makeatletter
\def\caption{\H@refstepcounter\@captype\@dblarg{\@caption\@captype}}
\makeatother

\title[Magnetic effects on NGC~5055 rotation]
{Magnetic contribution to outer-disk rotation in the spiral galaxy \ngc}
\author[N. Wo\'zniak et al.]{
Natalia Wo\'zniak,$^{1}$
Joanna Ja{\l}ocha,$^{2}$
{\L}ukasz Bratek$^{2}$
and Szymon Sikora$^{3}$\thanks{E-mail: szymon.sikora@uj.edu.pl}
\\
$^{1}$Doctoral School, University of the National Education Commission,
ul. Podchor\k{a}{\.z}ych 2, PL-30084 Krak\'ow, Poland\\
$^{2}$Institute of Physics, Cracow University of Technology,
ul. Podchor\k{a}{\.z}ych 1, PL-30084 Krak\'ow, Poland\\
$^{3}$Astronomical Observatory, Jagiellonian University, Orla 171,
PL-30244 Krak\'ow, Poland
}
\date{Accepted XXX. Received YYY; in original form ZZZ}
\pubyear{2026}
\begin{document}
\label{firstpage}
\pagerange{\pageref{firstpage}--\pageref{LastPage}}
\maketitle
\raggedbottom

\begin{abstract}
We test whether fields of galactic strength enhance NGC~5055's outer
gas rotation. Finite analytic disk decompositions yield consistent
rotation--density pairs and a blended gravitational reference approximately
retaining inner rotation while its density approaches the measured outer
H\,{\sc i}+He profile. This reference is modelled, not an independently
measured stellar curve. Its difference from fitted gas rotation gives
the required purely toroidal field for circular, axisymmetric midplane
motion. The prescribed field strength at the reconstruction's inner
radial boundary is $B_s$. With hydrogen inertia, no pressure and $B_s=0$,
constant exponential heights of 1~kpc and 250~pc give approximate peaks
of 5.47 and 10.95~$\mu$G near 34.9~kpc; 0.2~kpc gives approximately
12.24~$\mu$G. Extrapolated flaring exponential and Gaussian profiles
peak at approximately 5.25 and 5.96~$\mu$G, respectively, near 33.8~kpc.
For $B_s=8\mug$, their fields decline overall to approximately 5.73
and 6.30~$\mu$G near 39.7~kpc. 
These estimates are conditional
on the gravitational reference, vertical structure and prescribed boundary
field. Comparison with total radio fields supports the field scale
without establishing a magnetic origin for the rotation excess.
\end{abstract}

\begin{keywords}
galaxies: kinematics and dynamics -- galaxies: magnetic fields
\end{keywords}

\section{Introduction}

We investigate whether fields of strength comparable to those inferred
in spiral galaxies can increase gas rotation speed. Magnetic stresses
may also slow rotation or have little effect, depending on field strength,
configuration, radial gradient and gas density
\newcitep{Elstner2014,Khademi2023}.

The reported Milky Way stellar--gas rotation separation motivates this
investigation. \citet{Jalocha2016} found faster outer-disk gas rotation,
with the separation increasing with Galactocentric radius. Their magnetic
interpretation required fields of order $10\mug$, comparable to galactic
estimates.

Substantial measurement uncertainties and possible non-axisymmetric gas
motions leave other explanations open. Supernovae drive gas turbulence
\newcitep{Joung2006}, but explaining a systematic azimuthal excess requires
a mechanism reproducing its radial trend. The reported separation does
not uniquely identify its origin.

\citet{Nelson1988} and \citet{Battaner1992} discussed magnetic contributions
to outer rotation; \newcitet{Sanchez2004} and \newcitet{Sanchez2013}
subsequently emphasised radial confinement and boundary constraints.
An inward magnetic force under specified conditions must therefore be
distinguished from complete global gas equilibrium.

\newcitet{Jalocha2012a} and \newcitet{Jalocha2012b} applied related disk
inversions to NGC~891 and NGC~253, respectively. The former examined
rotation, vertical gradients and mass-to-light ratios, matched the outer
density to gas, and considered magnetic effects; the latter investigated
a magnetic contribution to starburst-galaxy rotation. We retain this
inverse approach, using finite analytic decompositions, \ngc's extended
\hi distribution and explicit tests of vertical-profile dependence.

For \ngc, we prescribe a gas rotation curve, gravitational reference
and gas density, then reconstruct the azimuthal field required by their
radial force residual. Without a separate outer-disk stellar--gas rotation
comparison, the gravitational reference is chosen within a specified
model class. Mass and field estimates are conditional on this choice,
uniquely determining neither stellar rotation nor the decomposition
of dark and luminous matter.

\subsection{Observational context and modelling limitations }
\label{sec:whythisgalaxy}

We test our hypothesis with \ngc (M63, the Sunflower Galaxy) in Canes
Venatici, adopting the approximately $7.2\Mpc$ distance of
\citet{Battaglia2006}. Its extended \hi disk constrains rotation and gas
surface density well beyond the bright stellar disk. Abundant gas and
dust motivate examining magnetic effects, expected to become increasingly
important in tenuous outer gas. Active star formation favours sustained
large-scale dynamo fields.

Under energy equipartition, \citet{Basu2013} inferred a mean total field
of approximately $9.5\mug$ over the radio-emitting disk, with about
$10.2\mug$ in arms and $9.9\mug$ between them. The relatively modest
central field, approximately $14.5\mug$, declines to about $8\mug$
near the mapped disk edge. These synchrotron estimates do not measure
the ordered azimuthal field at $R\simeq30$--$40\kpc$ reconstructed here.

Magnetic stresses may affect tenuous outer-gas rotation
\citep{Nelson1988,Battaner1992}, with magnitude and sign depending on
field geometry, radial gradients and gas density. \newcitet{Knapik2000}
detected a spiral magnetic pattern with a substantial radial component.
Our purely toroidal field explicitly simplifies the force calculation
rather than reproducing the measured geometry.

Despite considerable overall kinematic regularity, \ngc's extended
\hi disk has a pronounced warp and mild morphological and kinematic
lopsidedness. Its companion UGC~8313's dynamical influence remains
uncertain \citep{Battaglia2006}. These features limit strictly planar,
axisymmetric modelling and may affect the inferred outer force residual
and its attribution to gravitational or magnetic support.

\subsection{Observational inputs}
\label{sec:data}

We reconstructed rotation and \hi surface-density points from
Figs.~10 and 13 of \cite{Battaglia2006}, respectively, adopting
$7.2\Mpc$ for \ngc and multiplying the \hi surface density by $4/3$
to include helium.

We reconstructed the stellar luminosity profile from the major-axis
surface-brightness measurements in Table II of \cite{Kent1987},
taken as representative of the radial stellar distribution.
The disk luminosity surface density, in $L_\odot\,\mathrm{pc}^{-2}$,
was calculated using
$
\Sigma_L(R)=
\cos(\iota(R))\,
10^{-\frac{2}{5}
\left(
\mu(R)-M_{\odot,r}-M_\circ
\right)}$,
$
M_\circ=
5\log_{10}\!\left(\frac{180\cdot3600}{\pi}\right)-5$.
Kent's measurements are in the $r$ system of \citet{Thuan1976}.
We adopt the solar absolute magnitude $M_{\odot,r}=4.68$.
\citet{Kent1987} notes increased uncertainty in the outermost points
because \ngc filled the CCD field. The variable inclination profile
$\iota(R)$ follows \cite{Battaglia2006}; its luminosity-weighted effective
value for Kent's $\mu$ profile is approximately $62.4^\circ$.

\section{The smooth primary rotation curve}
\label{sec:rotcrv}

We model the rotation measurements with a single analytic curve using
the formal auxiliary mass function ${\cal M}(R)=R v_c^2(R)/G$.
For a flattened disk, ${\cal M}(R)$ represents rotation, not the mass
enclosed within $R$.

\begin{figure}
\centering
\includegraphics[width=\columnwidth]{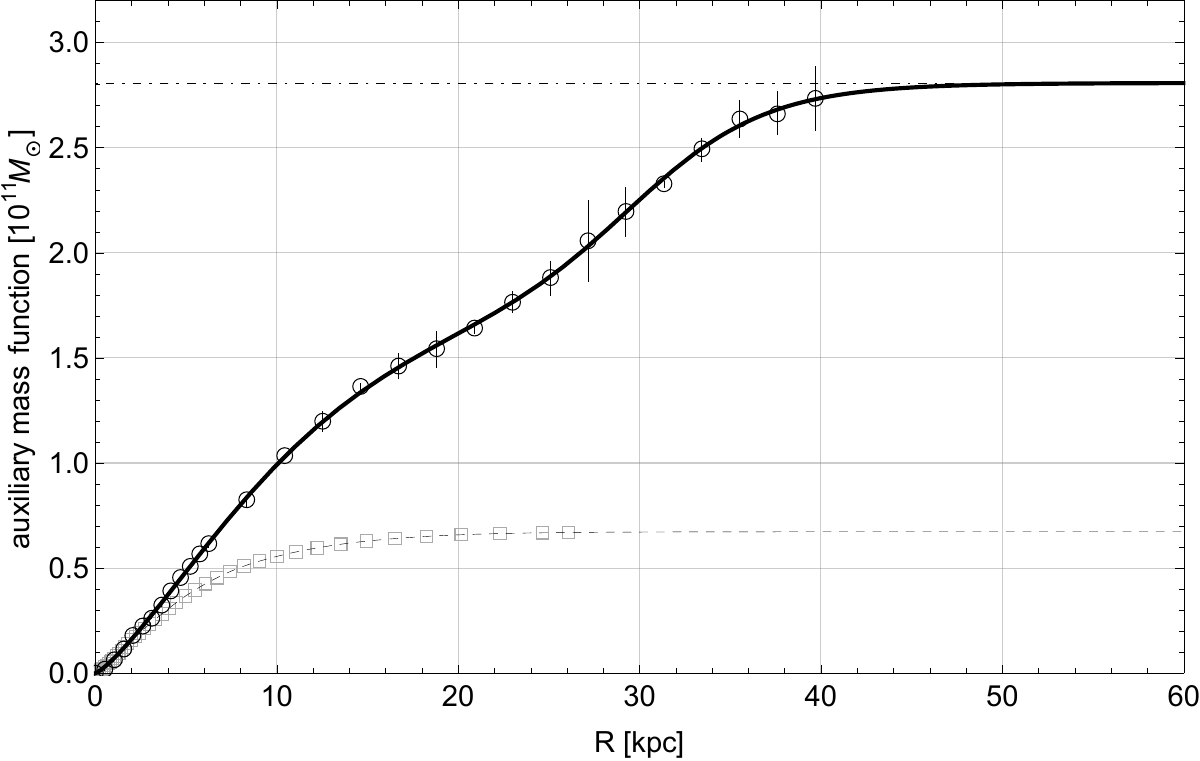}
\caption{
The smooth primary model $V_p(R)$ expressed as the auxiliary mass
function $\mathcal{M}_{\rm fit}(R)$ (solid line), asymptoting to
$2.81\times10^{11}\msun$. Circles with error bars show rotation
measurements. Dashed: cumulative luminosity scaled by
$\Upsilon\simeq4.83\mlsun$, giving a disk-integrated luminous mass
of approximately $6.73\times10^{10}\msun$. Squares mark luminosity
sample radii.}
\label{fig:masf}
\end{figure}

First fitting $R<22\kpc$ with
$\mathcal{M}_{\rm in}(R)=M\tanh^{\kappa}(R/a)$ gives
$\kappa=1.2436\cdots$. We then fit the full auxiliary mass function
with the smooth two-component model
\begin{equation}
\begin{split}
\mathcal{M}_{\rm fit}(R)={}&M_1\tanh^{\kappa}(R/a_1)\\
&+M_2\frac{\tanh((R-R_0)/a_2)+\tanh(R_0/a_2)}
{1+\tanh(R_0/a_2)}.
\end{split}
\end{equation}
The fitted parameters are $a_1=13.59(34)\kpc$,
$M_1=1.765(40)\times10^{11}\msun$, $a_2=7.33(55)\kpc$,
$M_2=1.043(57)\times10^{11}\msun$ and $R_0=29.75(24)\kpc$,
with asymptotic value $M_1+M_2=2.808(69)\times10^{11}\msun$.
\figref{fig:masf} shows $\mathcal{M}_{\rm fit}$.
The primary velocity model (\figref{fig:expl} and \figref{fig:rotc}) is
\begin{equation*}
V_p(R)\equiv[G\mathcal{M}_{\rm fit}(R)/R]^{1/2}.
\end{equation*}
Its Keplerian asymptote is adopted beyond the observed range,
rather than representing an observed rotation decline.

The luminous mass function in \figref{fig:masf} follows from interpolating
the logarithm of the luminosity surface density $\Sigma_L(r)$ within
its measured range and continuing it exponentially outside.
Integrating $2\pi r\Sigma_L(r)$ gives cumulative luminosity;
scaling by constant $\Upsilon$ gives $M_\star(R)=\Upsilon L(R)$.
A one-parameter fit to the inner auxiliary mass function yields
$\Upsilon\simeq4.83\mlsun$ and
$M_\star(\infty)\simeq6.73\times10^{10}\msun$.

The primary asymptotic mass parameter divided by the adopted total
luminosity is approximately
$\Upsilon(M_1+M_2)/M_{\star}\sim4.17\Upsilon\sim20.1\msun/\lsun$.
Including the fitted \hihe mass of approximately $8.3\times10^9\msun$
leaves approximately $73\%$ unaccounted for by stars and gas.
Identifying this specifically as dark matter requires additional
stellar-population and three-dimensional gravitational-geometry
assumptions.

\section{Toomre disk models and finite decompositions}

We use analytic disk velocity--density pairs from the Toomre family
\citep{Toomre1963}. A finite superposition defines both a surface density
and its squared circular velocity. The linear relation is between
surface density and squared velocity, not velocity itself. Exponential
and Gaussian disk families can be used in the same way.

\subsection{A formal definition}

Consider the recursive mapping
\begin{equation}
T_n(R)=-\frac{a^{2n-2}}{2n-3}\frac{\partial}{\partial a}
\br{\frac{T_{n-1}(R)}{a^{2n-3}}},\qquad n\geq2,
\end{equation}
where $T_n=\{\Sigma_n,M_n,(V_c^2)_n\}$ and the seed is
\begin{equation}
\begin{split}
\Sigma_1(u)&=\frac{M}{2\pi a^2}(1+u^2)^{-3/2}, \quad (V_c^2)_1(u)=\frac{GM}{a}\frac{u^2}{(1+u^2)^{3/2}}\\
M_1(u)&=M\br{1-(1+u^2)^{-1/2}}.
\label{eq:seed}
\end{split}
\end{equation}
Here $u=R/a$, $M$ is the total disk mass, and $M_n(R)$ is the
enclosed disk mass for the $n$th member.
With $R$ and $M$ fixed during differentiation, the recursion is
equivalently
\begin{equation}
\begin{split}
T_n&=D_{n-1}\circ\cdots\circ D_1(T_1),\qquad
D_k&=1-\frac{a\partial_a}{2k-1}.
\end{split}
\label{eq:operator}
\end{equation}

\subsection{Application to the present problem}
\label{sec:decomp}

Given characteristic radii $R_{m,k}$ ($k=0,\dots,K$), for example
geometrically spaced over $a\le R\le b$ with $a=R_{m,0}$ and $b=R_{m,K}$,
we approximate realistic disk profiles by $\sum_k C_k T_{m,k}(R)$.
Least-squares fitting determines $C_k$ through a linear system, with
$|C_k|$ analogous to squared velocity scales $V_{m,k}^2$. As formal
expansion amplitudes, these coefficients may be negative; individual
contributions need not represent physically admissible mass distributions.
Their superposition simultaneously defines the best-fit squared rotation
curve and associated surface-density and mass-function profiles.

At finite accuracy, decompositions need not be unique, and inferred
profiles can depend on the basis $T_{m,k}$ and its continuation. We used this method
to construct the realistic rotation-curve--surface-density pairs presented
here, employing analogous expansions in exponential and Gaussian disk
families in some cases. The coefficients $C_k$ are typically strongly correlated, alternate in sign
and may greatly exceed the characteristic squared velocity scale.
Sufficient numerical precision nevertheless allows reconstruction to
the required accuracy while maintaining a positive total surface density.

For a given $T_{m,k}$ we identify the characteristic radius  with the position 
of the corresponding rotation-curve maximum $(R_m,V_m)$. As an illustration, consider the $T_2$ element
\begin{equation}
\begin{split}
\Sigma_2(R)&{=}\frac{3M}{2\pi a^2}(1+u^2)^{-5/2}, \quad
(V_c^2)_2(R){=}\frac{GM}{a}
\frac{u^2(u^2+4)}{(1+u^2)^{5/2}}.
\label{eq:toomre2}
\end{split}
\end{equation}
With $R_m$ measured
in $\kpc$ and $V_m$ in $\km/\sec$, this becomes
{$\Sigma_2(R)=0.118768\cdots V_m^2 R_m^{-1}/(1 + \alpha_2 x^2)^{5/2}$
in $\msun/\pc^2$} and
{$V_c^2(R)=4.05746\cdots V_m^2x^2 (1 + \alpha_2 x^2/4)
 /(1+\alpha_2 x^2 )^{5/2}$ in $(\km/\sec)^2$},
with total mass
$M=0.276701\cdots\times10^6 R_m V_m^2$ in $\msun$.
Here, $x=R/R_m$, and the maximum occurs at
$u_m^2=2\sqrt{6}-4 =0.898979\cdots\equiv \alpha_2$
(the other numerical coefficients will depend slightly on the adopted dimensional
constants).

\section{The smooth secondary rotation curve with decreased outer branch}
\label{sec:rotcrvII}

Stars move predominantly under gravity, whereas partially ionized gas
also interacts with the magnetic field. Magnetic forces may therefore
separate stellar and gas rotation curves. The systematically faster gas
rotation reported in the Milky Way was shown to be compatible with the
observed galactic magnetic field strength \cite{Jalocha2016}.
For external galaxies, usually only one rotation curve is available,
representing the mean observed kinematics. Separate stellar and gas curves
are measurable in principle, but weak signals often prevent their
recovery, or observations trace only gas.

\medskip

For \ngc, separate stellar and gas curves are unavailable. We therefore
lower the observed outer rotation numerically to construct a secondary
curve representing a possible magnetic contribution to gas rotation.

\begin{figure}
\centering
\includegraphics[width=0.95\linewidth]{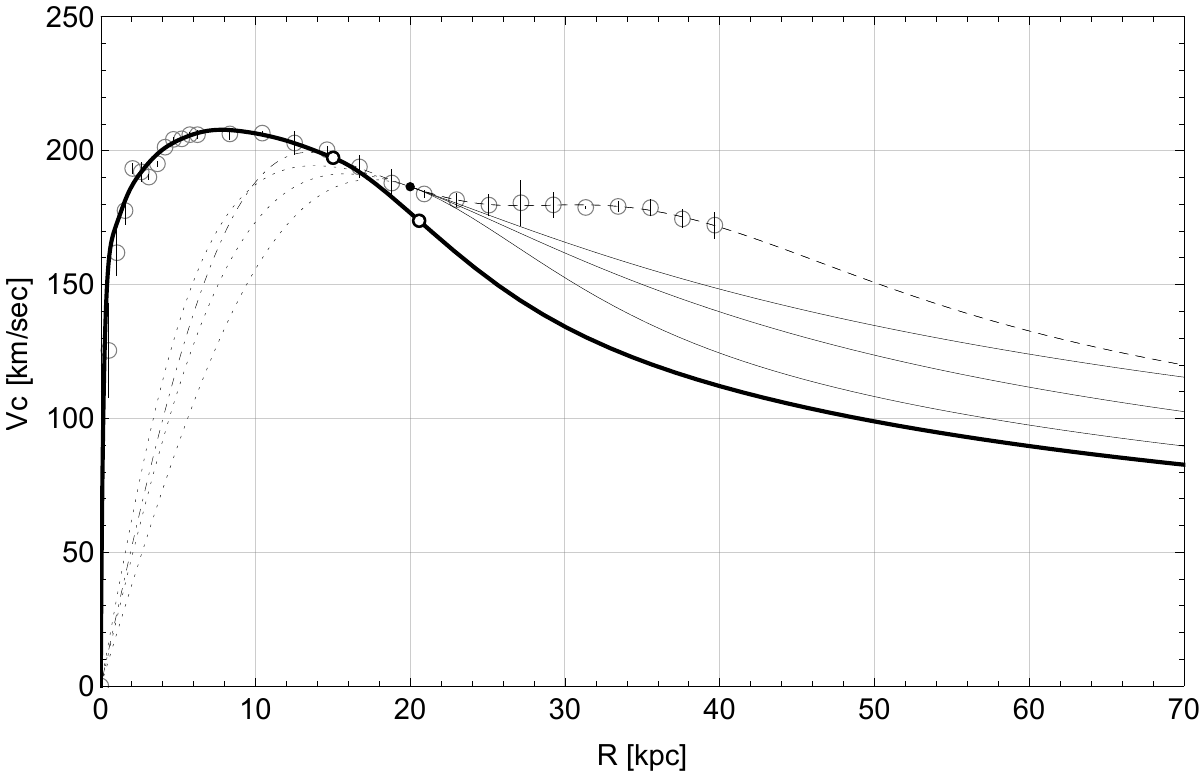}
\caption{\label{fig:expl}
{Three continuously differentiable models (thin solid) follow the
best-fitting primary (dashed) to an arbitrary join at $20\kpc$
(black dot), then tangent Toomre branches 1, 2 and 28 (inner branches
dotted). The adopted secondary (thick black solid) coincides with the
inner primary, then follows $T_{17}$ beyond the primary inflection
at $R_s\sim15.021\kpc$ (black circle with white centre).
The inner $T_{17}$ branch is dot-dashed; the secondary inflection
at $\sim20.580\kpc$ is marked. Gray circles and vertical error bars
show \ngc measurements and uncertainties from
\protect\cite{Battaglia2006}.}
}
\end{figure}

The local mass-to-light ratio rises rapidly beyond ordinary stellar
values, exceeding the global mean of approximately $20.1\msun/\lsun$
at $10$--$15\kpc$. We therefore modify the primary rotation-curve model of
\secref{sec:rotcrv} beyond this region. \figref{fig:expl} shows three
illustrative smooth continuations and the adopted secondary model.

Tangentially attaching the velocity component of a velocity--density
pair at a joining radius ensures continuity of the rotation curve and
its first derivative. The joining radius must balance preservation of
the well-constrained inner structure against substantial reduction of the
outer asymptotic mass: joining too far inward distorts the inferred inner
mass distribution, whereas joining too far outward makes the reduction
inefficient.

Since the radius where magnetic effects become important is unknown
{\it a priori}, we choose a point intrinsic to the primary curve: its
distinguished inflection within the extended flattening region,
$R_s\sim15.021\kpc$ and $V_s\sim197.30\km/\sec$. This fixes the joining
radius without an additional arbitrary parameter. The inflection marks
a change of curvature towards the outer flattened branch, where magnetic
support may become relevant. This motivates the choice without determining
where magnetic forces actually become important.

The secondary curve follows the primary for $R<R_s$ and a tangent
Toomre branch $T_{17}$ for $R>R_s$, with maximum
$R_m\sim 13.376\kpc$ and $V_m=199.30\km/\sec$.
The join has a continuous first derivative; the outer branch has an
inflection at $(20.580\kpc, 173.78\km/\sec)$ and an asymptotic mass of
approximately $1.0908\times 10^{11}\msun$, or $38.8\%$ of the primary
value. The tangent Toomre curve is also dot-dashed in \figref{fig:rotc},
showing the small difference between the secondary and final blended
model of \secref{sec:blended}.
Because thin-disk rotation and density are related nonlocally, modifying
the outer curve also redistributes the inferred inner density. Despite
the substantial asymptotic mass reduction, this redistribution remains
modest relative to the primary model.

\section{The final blended density-velocity pair}\label{sec:blended}

We further reshape the secondary rotation curve (\secref{sec:rotcrvII})
to match the observed outer gas content while preserving the original
inner rotation structure. This tests the hypothesis that magnetic
interaction with gas accounts entirely for the excess outer rotation
represented by the primary model.

We first expand the independently measured \hihe surface density
$\Sigma_{\rm g}(R)$ in a finite basis (\secref{sec:decomp}), calculate
its rotational contribution $V_{\rm g}(R)$ and subtract it from the
secondary model:
$V_{\rm r}^2(R)=V_{\rm sec}^2(R)-V_{\rm g}^2(R)$.
We then expand this reduced curve in a finite basis. Increasing the
number of adjustable parameters permits essentially arbitrary numerical
accuracy over the fitted interval, while the choice and distribution
of basis maxima leave some freedom in the outer density tail.
We use this freedom to regularise the density physically, selecting a
decomposition whose reconstructed density $\Sigma_{\rm r}(R)$ remains
positive everywhere but falls well below the gas density at large radii.
This reflects the expectation that the luminous stellar component
becomes negligible in the outer gas-dominated disk. Restoring the gas
gives the blended surface density
$\Sigma_{\rm blend}(R)=\Sigma_{\rm r}(R)+\Sigma_{\rm g}(R)$.

The subtraction, regularisation and restoration introduce only negligible
inner density changes, preserving the secondary model's inner rotation.
At large radii, the non-gaseous density becomes negligible and the blended
density approaches the observed \hihe profile.

\begin{figure}
    \centering
    \includegraphics[width=0.95\linewidth]{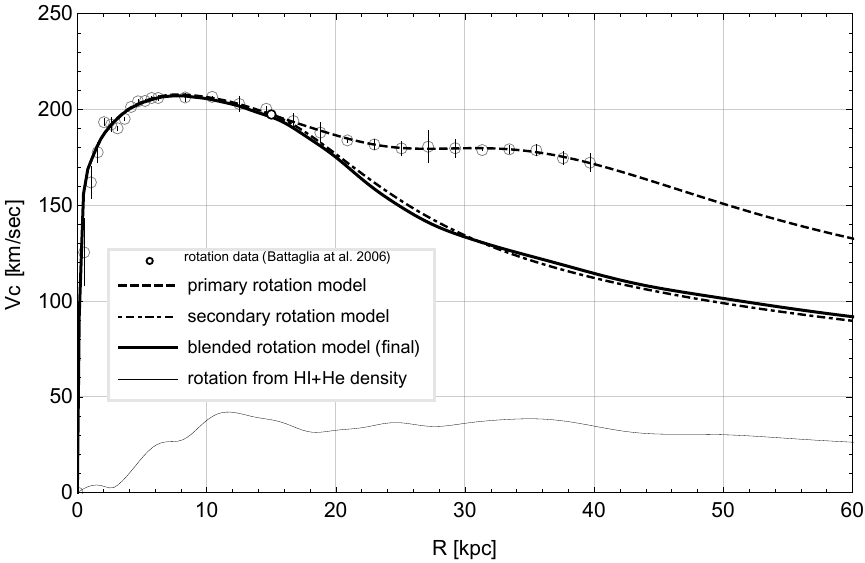}
    \caption{\label{fig:rotc}
{Rotation models for \ngc. Open gray circles and vertical error bars:
measurements and uncertainties from \protect\cite{Battaglia2006};
dashed: fitted primary. The secondary (thin dot-dashed) smoothly joins
a Toomre $T_{17}$ branch at the primary inflection,
$R_s\sim15.021\kpc$, $V_s\sim197.30\km/\sec$
(black circle with white centre); the branch maximum is
$R_m\sim13.376\kpc$, $V_m=199.30\km/\sec$.
The adopted blended gravitational curve (thick black solid) differs
from the secondary only by small smooth oscillations from nonlocal
mass redistribution, indistinguishable inside at this resolution.
Its density smoothly approaches the observed \hihe density where
measured stellar luminosity is negligible. Thin solid: rotation
from the observed \hihe alone.}
}
\end{figure}

The decomposition amplitudes uniquely determine the corresponding
blended rotation curve. Thin-disk gravity is strongly nonlocal, so density
redistributions introduce small smooth oscillations in the rotation
curve. Nevertheless, the blended curve remains nearly indistinguishable
from the secondary model, including its outer $T_{17}$ branch
(\figref{fig:rotc}). The blending therefore modifies mainly the inferred
density, yielding a consistent rotation--density pair that reproduces
the inner observed rotation and outer gas distribution. We adopt this
curve as the final gravitational reference for estimating the magnetic
contribution to gas rotation.

\begin{figure}
    \centering
    \includegraphics[width=0.95\linewidth]{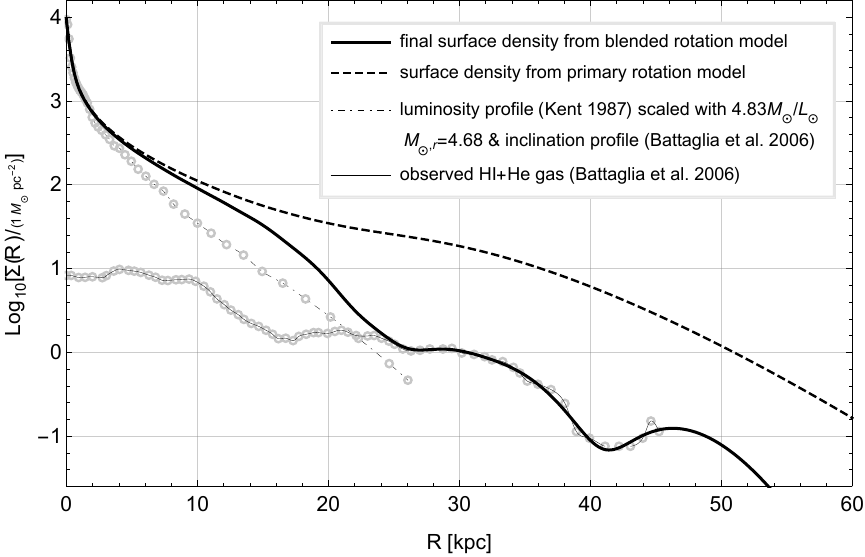}
    \caption{\label{fig:gr-dens}
{Disk surface mass densities for \ngc, in $\msun/\pc^2$ on a
$\log_{10}$ scale. Black dashed: apparent density from the smooth,
asymptotically Keplerian primary model of observed rotation.
Thick black solid: {blended} gravitating density, smoothly matching
the outermost gas density. Thin gray interpolants show observed \hihe
density (solid) and local brightness (dot-dashed), the latter scaled
by the optimum mass-to-light ratio $\sim4.83\,\msun/\lsun$ to fit
the innermost apparent density. Circles mark gas and brightness
measurements.}
}
\end{figure}

The density comparison (\figref{fig:gr-dens}) shows the main effect.
The primary model remains well above the observed outer \hihe density,
requiring substantial additional gravitating matter. The blended density
instead declines rapidly beyond the luminous disk and approaches the gas
density, so the local outer gravitating surface density is approximately
accounted for by the observed gas. The luminosity
profile, scaled by the optimum mass-to-light ratio, reasonably reproduces
the inner density but departs from both gravitating profiles at large
radii, highlighting the outer luminous--dynamical mismatch in the absence
of magnetic corrections.

\subsection{Local mass-to-light ratio}

\begin{figure}
    \centering
    \includegraphics[width=0.95\linewidth]{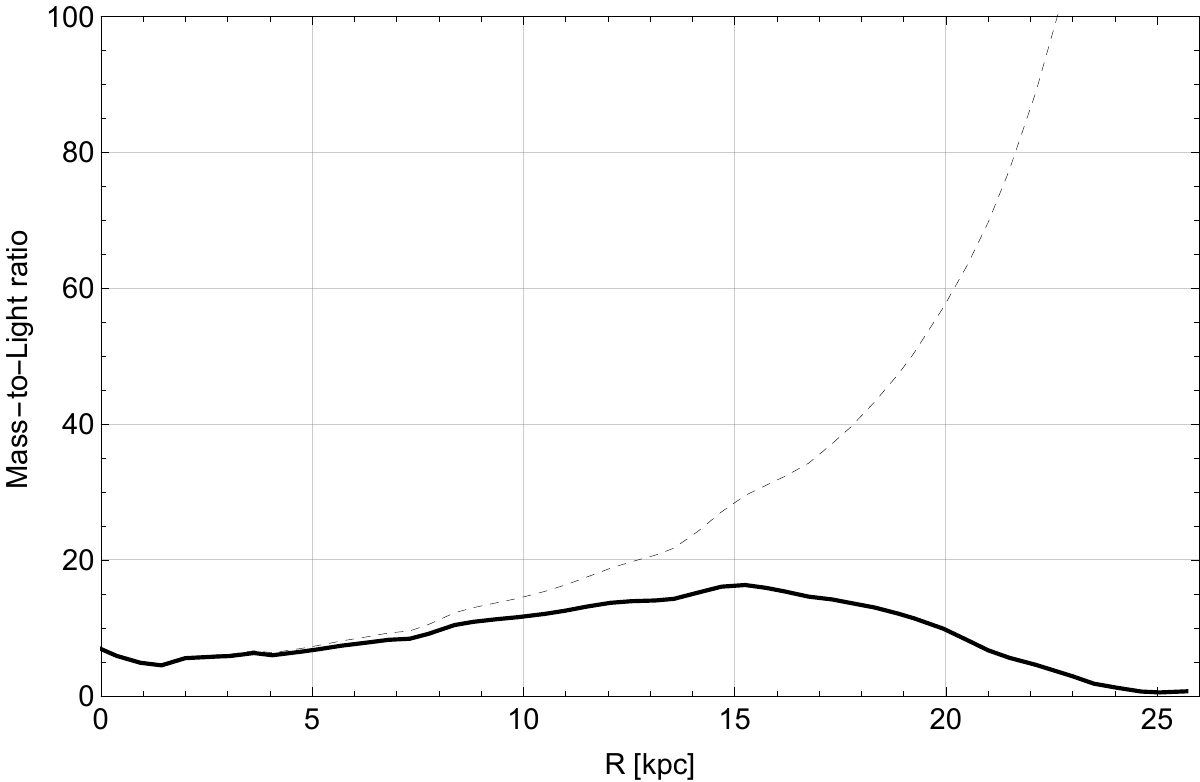}
    \caption{\label{fig:gr-lumi}
{Gas-subtracted local mass-to-light ratios for \ngc. Black dashed: primary
model of observed rotation with its asymptotically Keplerian extension.
Black solid: adopted blended gravitating-density model for estimating
magnetic effects, whose density smoothly matches the observed gas
at the outermost radii.}    
}
\end{figure}

\figref{fig:gr-lumi} shows the gas-subtracted local mass-to-light ratios,
$(\Sigma-\Sigma_{\rm g})/\Sigma_L$, for the primary and blended models.
They include any inferred non-gas gravitating matter, not only stars.
For the primary model,
the ratio rises rapidly to very large outer values: the required
gravitating density increasingly exceeds that traced by light,
conventionally indicating substantial non-baryonic dark matter.

The blended model behaves differently. Through the nonlinear
velocity--mass relation, a modest rotation reduction substantially changes
the inferred density, especially at large radii. Its mass-to-light ratio
reaches a broad maximum near $R\sim15\kpc$ and then approaches zero as
the blended density approaches the gas density. This constructed decrease
illustrates how a magnetic contribution could reduce the non-gas
gravitating density inferred from the observed rotation.

\section{Inclusion of magnetic interaction with partially ionized gas}
\label{sec:magnetic}

Spiral-galaxy rotation curves, including that of \ngc, usually trace disk
gas, primarily neutral hydrogen. Partial ionization allows interaction
with the large-scale magnetic field, potentially increasing gas rotation
above the gravity-only value. 
The relatively strong field of the large spiral galaxy \ngc motivates
testing its dynamical influence. We estimate the field required to
account for the difference between the observed and gravitational
rotation curves in \figref{fig:rotc}.

\subsection{Vertical gas structure}
\label{sec:height}

We reconstruct the \hi volume density from the measured surface density
using a normalised exponential profile with scale height $h(R)$,
truncated at $\Delta=3h$:
\begin{equation}
\rho_{\rm H}(R,z)=
\frac{\sigma_{\rm HI}(R)e^{-|z|/h(R)}}
{2h(R)(1-e^{-3})},\qquad |z|<3h(R),
\label{powNaObj}
\end{equation}
and zero outside this layer. Here $\sigma_{\rm HI}=(3/4)\Sigma_{\rm g}$
is the hydrogen surface density. The exponential half-width at
half-maximum (HWHM) is $H=h\ln2$.

For constant exponential scale height, we use $h=1\kpc$ as the reference
case and compare its $B_s=0$ separatrix with those for  $750$, $500$ and also  $h=250\mathrm{pc}$ close
to $h=0.2\kpc$ adopted for the gaseous disk in \citet{Battaglia2006}.
The reference case will also show the family of solutions with different boundary fields. These heights
are adopted comparison parameters. The surface-density input traces total \hi and
does not distinguish cold and warm neutral phases.

To test flaring, we adopt the linear fit to the \ngc HWHM profile
in fig.~8 of \newcitet{Patra2020}:
$H(R)=\sq{91.5+24.6(R/\mathrm{kpc})}\mathrm{pc}$.
This HWHM is derived from numerical hydrostatic profiles that are
approximately Gaussian. We use it to construct exponential and Gaussian
profiles with equal hydrogen surface density and HWHM, without reproducing
the full numerical solutions. Patra adopts $10.1\Mpc$, compared with
our $7.2\Mpc$, and includes a dark halo in the hydrostatic potential.
We retain the published coefficients to test the dependence of our
magnetic-field estimates on gas thickness, without recalculating vertical
force balance in our adopted gravitational potential. Beyond the
published radial range, approximately $17\kpc$, we extrapolate the fit,  out to 
$40\kpc$ where \hi rotation is still measured.

For the flaring exponential, the e-folding height is
\begin{equation}
h(R)=\frac{H(R)}{\ln2}=\sq{0.1320+0.03549(R/\mathrm{kpc})}\,\mathrm{kpc}.
\label{eq:flaring-explicit}
\end{equation}
It is approximately $0.664$, $0.842$, $1.197$ and $1.552\kpc$ at
$R=15$, $20$, $30$ and $40\kpc$, respectively. The Gaussian comparison
profile, motivated by Patra's approximately Gaussian solutions, is
\begin{equation}
\rho_{\rm H}(R,z)=
\frac{\sigma_{\rm HI}(R)\sqrt{\ln2}}{\sqrt{\pi}H(R)}
\exp\br{-\ln2\frac{z^2}{H^2(R)}}.
\label{eq:gaussian}
\end{equation}
It has HWHM $H(R)$ and is normalised over all $z$, whereas the exponential
retains the original $3h$ truncation.

Throughout these comparisons, the primary and blended rotation curves
and gas surface density remain fixed. The gravitational calculation
retains its razor-thin disk kernel: varying the height changes the
midplane inertia in the magnetic inverse, without refitting the full
three-dimensional gravitational potential.

For comparison with the original field calculation, the numerical values
below use $\rho=\rho_{\rm H}(R,0)$. The gravitational gas density already
includes helium through our retained $4/3$ convention;
\citet{Battaglia2006} instead uses $1.4$. This compositional correction
is independent of the vertical truncation normalisation  (however, note the coincidence $(4/3)/(1-e^{-3})\approx 1.4$).

If helium shares the hydrogen motion and magnetic coupling, the inertial
density becomes $(4/3)\rho_{\rm H}$. With unchanged velocities, pressure
prescription and zero boundary field, all quoted field amplitudes then
increase by $\sqrt{4/3}\simeq1.155$. This is an inertia-convention effect,
distinct from the dependence on vertical structure.

\subsection{Magnetic field geometry approximation}

We consider the large-scale coherent field responsible for the averaged
magnetic stress in radial force balance, rather than small-scale turbulent
fluctuations. Observed spiral-galaxy fields generally follow spiral
structures and are often predominantly azimuthal within the thin disk
\cite{Beck2015}. Differential rotation, galactic dynamo action and shear
amplification also favour predominantly toroidal configurations. Thus,
$\mathbf B\approx B_\phi(R,Z)\,\mathbf e_\phi$
is a physically motivated, standard approximation in axisymmetric
large-scale galactic MHD models.

The azimuthal field contributes to radial force balance through magnetic
tension and pressure. We neglect radial and vertical components as a
modelling simplification, without assuming their stresses are small. Partially ionized
H\,{\sc i} gas may remain coupled to this field, whereas effectively
collisionless stars are largely insensitive to magnetic forces.
For the $1\kpc$ and flaring models, inferred strengths of order
$\sim5\mug$ are observationally realistic for galaxies.

For rotation near the disk mid-plane, we adopt the axisymmetric
approximation
$\mathbf B\simeq B_\phi(R)\,\mathbf e_\phi$.
Neglecting explicit $Z$-dependence defines a mid-plane approximation
for estimating the toroidal magnetic contribution to \hi rotation,
rather than a complete three-dimensional field description.

For stationary ideal MHD with cylindrical symmetry,
infinite electrical conductivity, and negligible dissipation and heat
conductivity, we parametrise the field by its inclination tangent
$T_{\iota}(R)$ and amplitude $B(R)$:
$B_R(R)-T_{\iota}(R)B_{\phi}(R)=0$ and
$B_R^2(R)+B_{\phi}^2(R)=B^2(R)$. The radial Euler equation then gives
\begin{eqnarray}
\frac{B^2(R)}{8\pi}+\frac{C_{\upsilon}^2}{R^2\rho_{\rm gas}(R)}=
\frac{1}{R^2}\int
\Big( V^2_{\phi,{\rm gas}}(R)-V^2_{\phi,{\rm grav}}(R) \Big.
\nonumber \\
-\Big. \frac{R p'_{\rm gas}(R)}{\rho_{\rm gas}(R)}
+\frac{C_{\upsilon}^2}{R^2\rho_{\rm gas}^2(R)} \Big)
\rho_{\rm gas}(R) R\ud{R}\nonumber
\end{eqnarray}
consistent with mass continuity and a divergence-free field:
$$
T_{\iota}(R)=\frac{C_{\iota}}{\sqrt{R^2B^2(R)-C_{\iota}^2}},\qquad
V_{R,{\rm gas}}(R)=\frac{C_{\upsilon}}{R\,\rho_{\rm gas}(R)},
$$
with three integration constants: $C_{\iota}$, $C_{\upsilon}$, and the
implicit constant in the free term $C/R^2$ from indefinite integration.
The azimuthal Euler equation additionally requires
$\frac{R B(R)C_{\iota}}{(4 \pi ) \sqrt{R^2 B(R)^2-C_{\iota}^2}}
{\partial_R (R B(R))}=C_{\upsilon}
{\partial_R (R V_{\phi,{\rm gas}}(R))}$,
which integrates to
$$
C_{\iota} \sqrt{-C^2_{\iota}+R^2 B(R)^2}
=4\pi C_{\upsilon}  R V_{\phi,{\rm gas}}(R)+C_{\phi}
$$
with another integration constant $C_{\phi}$. For nonzero inclination,
the two integral expressions are mutually consistent only in special
cases.

We adopt circular field lines, $C_{\iota}=0$, consistent with arbitrary
pure rotation $V_{\phi,{\rm gas}}(R)$, $V_{{R},{\rm gas}}(R)=0$.
Writing the inner boundary as $R_o=R_s$ and its field as $B_o=B_s$,
the resulting expression is
\begin{align}\label{eq:MHD}
B^2(R)=\frac{R_o^2}{R^2}B_o^2  \\
+\frac{8\pi}{R^2}\int\limits_{R_o}^{R}\rho_{\rm gas}(r)
\br{V^2_{\phi,{\rm gas}}(r)-V^2_{\phi,{\rm grav}}(r)-
\frac{r\, p'_{\rm gas}(r)}{\rho_{\rm gas}(r)}} r\ud{r}.\nonumber
\end{align}
The pressure may be taken as isothermal, $p(R)=c^2_s\rho(R)$, where
$c_s$ is the isothermal sound speed, as in \cite{Nelson1980}.
We neglect its contribution in the adopted model.

\subsection{Can the pressure term be safely ignored?}

The effective H\,{\sc i} sound speed should reflect the neutral gas
phase. For thermal hydrogen, up to factors of order unity,
$c_s\simeq (kT/m_{\rm H})^{1/2}$,
giving $c_s\sim1\,{\rm km\,s^{-1}}$ for the cold neutral medium,
$T\sim10^2\,{\rm K}$, and $c_s\sim8$--$10\,{\rm km\,s^{-1}}$
for the warm medium, $T\sim10^4\,{\rm K}$. Values from several
${\rm km\,s^{-1}}$ to roughly $10\,{\rm km\,s^{-1}}$ are therefore
natural for H\,{\sc i} discs, particularly when $c_s$ includes unresolved
turbulent broadening. For a purely thermal interpretation, significantly
larger values would correspond to hot ionized rather than ordinary
neutral H\,{\sc i} gas.

For an illustrative pressure estimate in \eqref{eq:MHD}, we adopt an
exponential radial density with scale length ${d}\simeq5$--$10\kpc$.
For an isothermal equation of state,
$p_{\rm gas}=c_s^2\rho_{\rm gas}$,
with approximately constant $c_s$, the integrand's pressure term is
$
-\frac{R\,p'_{\rm gas}(R)}{\rho_{\rm gas}(R)}
=
-c_s^2\,R\,\frac{\rho'_{\rm gas}(R)}{\rho_{\rm gas}(R)}.
$
For
$
\rho_{\rm gas}(R)\propto \exp(-R/d),
$
this becomes
$$
-\frac{R\,p'_{\rm gas}(R)}{\rho_{\rm gas}(R)}
=
c_s^2\frac{R}{d}.
$$
With the standard H\,{\sc i} value $c_s\simeq8\,{\rm km\,s^{-1}}$,
at $R\simeq30$--$40\kpc$ we obtain
$
-\frac{R\,p'_{\rm gas}(R)}{\rho_{\rm gas}(R)}
\sim
200\text{--}500\,{\rm km^2\,s^{-2}}
$.
We compare this with the rotational residual
$
V_{\phi,{\rm gas}}^2-V_{\phi,{\rm grav}}^2
$.
For the adopted rotation models, the difference
$
V_{\phi,{\rm gas}}-V_{\phi,{\rm grav}}
$
rises from approximately $0\,{\rm km\,s^{-1}}$ at the splitting point
to $45\,{\rm km\,s^{-1}}$ at $30\kpc$ and
$60\,{\rm km\,s^{-1}}$ at $40\kpc$. Assuming
$
V_{\phi,{\rm gas}}+V_{\phi,{\rm grav}}
\sim400\,{\rm km\,s^{-1}},
$
gives
$$
V_{\phi,{\rm gas}}^2-V_{\phi,{\rm grav}}^2
\sim10^4\,{\rm km^2\,s^{-2}},
$$
exceeding the illustrative pressure estimate by more than an order of
magnitude. For the adopted constant-height density fit, the local pressure
correction is approximately $6$ per cent near the last rotation
measurement. It remains subdominant there but may matter where the
rotational excess becomes small.

For steeper profiles, ${d}\simeq2$--$3\kpc$, pressure increases as
$1/d$, potentially reaching
$10^3\,{\rm km^2\,s^{-2}}$
at $R\simeq30$--$40\kpc$: appreciable, but still subdominant to the
rotational residual. Beyond \ngc's optical disc, the surface density
declines by only $\sim0.06\,M_\odot\,{\rm pc^{-2}\,kpc^{-1}}$
in the warped outer region \cite{Battaglia2006}; this absolute slope
does not determine an exponential scale length. \citet{Kalberla2008}
give a \hi surface-density scale of $3.75\kpc$ over
$12.5$--$30\kpc$ in the Milky Way. The pressure term here depends
on the midplane-density gradient, including any height gradient.

\subsection{Reconstructed magnetic field and thickness dependence}

\begin{figure}
\centering
\includegraphics[width=\columnwidth]{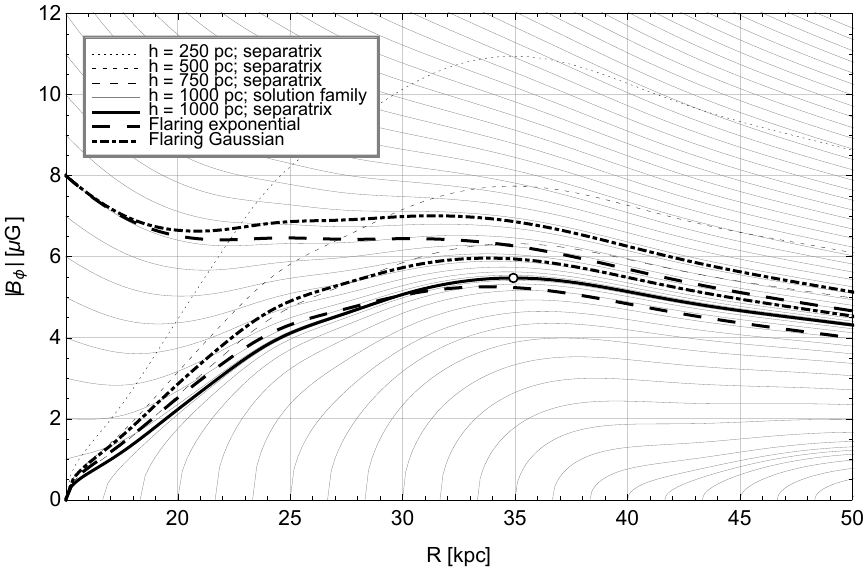}
\caption{Integral curves of the azimuthal magnetic field for the final
blended rotation curve. Thin grey solid curves form the solution
families for constant exponential scale height $h=1\kpc$; their $B_s=0$
separatrix is thick black, with its maximum, approximately $5.47\mug$ at
$34.9\kpc$, marked by an open circle. Thin black dotted, short-dashed and
medium-dashed curves show corresponding $B_s=0$ separatrices for $h=250$,
$500$ and $750\mathrm{pc}$, respectively. Thick black long-dashed and
dot-dashed curves show fields for flaring exponential and Gaussian profiles,
respectively, for $B_s=0$ and $B_s=8\mug$, extrapolated beyond the original
boundary. All height comparisons use hydrogen column density
and neglect pressure. Curves extend to $50\kpc$, continuing the model beyond
the last rotation measurement, near $39.7\kpc$.}
\label{fig:gr-mgnt}
\end{figure}

To test whether magnetic support could explain the proposed outer
rotation-curve reduction, we reconstruct the azimuthal field required
by the difference between observed gas rotation and the final blended
gravitational reference, used for all curves in \figref{fig:gr-mgnt}.
All quoted maxima assume hydrogen inertia, neglect pressure and use
$B_s=0$.

For constant exponential height $1\kpc$, the thick black separatrix
has a broad maximum of approximately $5.47\mug$ at $34.9\kpc$
(open circle); thin grey curves show the corresponding solution families.
With fixed radial inputs, no pressure and $B_s=0$, these models satisfy
$B_\phi\propto h^{-1/2}$, demonstrating appreciable thickness dependence.
The separatrices for $h=750$, $500$ and $250\mathrm{pc}$ peak at
approximately $6.32$, $7.74$ and $10.95\mug$, respectively, at the same
radius. An exponential height of $0.2\kpc$ would give approximately
$12.24\mug$. This unplotted case uses the same numerical height as
the gas-disk parameter of \citet{Battaglia2006}.

The flaring exponential and Gaussian profiles, using the extrapolated
HWHM prescription of \secref{sec:height}, peak at approximately $5.25$
and $5.96\mug$, respectively, near $33.8\kpc$. Their equal hydrogen
column densities and HWHM but different midplane densities yield
different required fields. Variable height must enter the radial
integral through the density; pointwise rescaling of a constant-height
field is insufficient.

Figure~\ref{fig:gr-mgnt} also shows flaring solutions with $B_s=8\mug$.
The non-zero boundary field increases the amplitude through the
integration constant without changing the radial force residual.
Beyond the last rotation measurement, near $39.7\kpc$, all curves
are model continuations.

These amplitudes are comparable to the equipartition total fields
inferred for \ngc: approximately $9.5\mug$ on average and $10.2\mug$
in spiral arms \citep{Basu2013} (\secref{sec:whythisgalaxy}).
The $1\kpc$ and flaring models require fields below the disk average;
thinner constant-height models require stronger fields. Under the
adopted assumptions, galactic-strength fields can therefore supply
the prescribed additional radial support. However, radio estimates
describe the total field in the brighter disk, not the ordered
azimuthal field at $30$--$40\kpc$. Amplitude agreement establishes
neither a magnetic origin for the excess rotation nor complete
magnetohydrodynamic equilibrium. Testing this interpretation requires
constraints on the outer field strength, radial gradient and geometry.

\section{Relation to previous work}
\label{sec:discussion}

Magnetic contributions to gas rotation depend on competing tension
and magnetic-pressure gradients. For circular motion and a purely
toroidal field, $\mathbf{B}=B_\phi(R)\mathbf{e}_\phi$, axisymmetric
midplane force balance in Gaussian cgs units gives
\begin{equation}
v_p^2-v_{\rm grav}^2=
\frac{R p'}{\rho}
+\frac{1}{8\pi\rho R}\frac{\ud{}}{\ud{R}}(R^2B_\phi^2).
\label{eq:radial}
\end{equation}
Here $v_p$ is the gas rotation speed, $v_{\rm grav}$ the gravitational
circular speed, and $p'=\ud{p}/\ud{R}$. The magnetic contribution is
\begin{equation}
v_{\rm mag}^2=\frac{B_\phi^2}{4\pi\rho}
\left(1+\frac{\ud{\ln B_\phi}}{\ud{\ln R}}\right).
\label{eq:sign}
\end{equation}
Despite its notation, $v_{\rm mag}^2$ is signed. A field decreasing
more slowly than $1/R$ exerts an inward force, increasing the rotation
speed required for radial balance; a faster decrease produces an
outward force and reduces it. Axisymmetry alone does not fix the sign.

Our approach reconstructs rather than prescribes the field. Over the
outer interval considered, we place the blended gravitational reference
below the primary gas rotation curve and reconstruct the field required
by their positive velocity-squared difference. Neglecting pressure,
this construction selects an inward magnetic force; we then assess
the plausibility of the required field strength and radial profile.

By contrast, \newcitet{Elstner2014} evaluated rotational effects of
synchrotron-constrained field profiles and examined dynamo simulations.
For a purely toroidal exponential field with radial scale length $L_B$,
the bracket in \eqref{eq:sign} becomes $1-R/L_B$: tension dominates
inside $L_B$, outward magnetic pressure outside. Their outer slowing
and our enhancement follow the same force balance but different radial
profiles and starting assumptions. Exponential radial profiles are
compatible with axisymmetry.

\newcitet{Khademi2023} improved rotation fits for NGC~6946 by including
three-dimensional magnetic geometry alongside dark-halo models,
illustrating the influence of spatial field structure. In \ngc,
\newcitet{Knapik2000} detected a spiral magnetic pattern with a
substantial radial component; our toroidal approximation isolates
a simpler geometry for directly reconstructing the required radial stress.

Other stresses and boundary conditions also distinguish these studies.
\newcitet{Sanchez2013} included ordered and tangled fields with thermal,
turbulent and cosmic-ray pressure, finding little net rotational change
in observationally constrained Galactic models. \newcitet{Sanchez2004}
found that radial confinement and global boundary conditions prevent
magnetic forces from replacing dark haloes in their models. These bounds
also apply to inverse solutions. For our archived gas and gravity
continuations, the $h=1\kpc$, $B_s=0$ reconstruction fails the necessary
confinement condition for an isolated, pressure-free toroidal exterior.
Our inverse therefore determines the stress required by a selected
rotation--density pair over a finite interval, without establishing a
globally confined equilibrium.

The boundary term $R_s^2B_s^2/R^2$ in \eqref{eq:MHD} supplies additional
freedom, changing the field amplitude without changing the prescribed
radial force residual. We consider $B_s=0$ separatrices and flaring
solutions with $B_s=8\mug$. These profiles quantify possible magnetic
contributions to the chosen rotation excess, testable through outer-disk
measurements of field strength, radial gradient and geometry.

\section{Conclusions}

We investigated whether large-scale fields of galactic strength could
contribute to \ngc's outer gas rotation. Finite analytic disk
decompositions yielded mutually consistent rotation curves and surface
densities, first for a secondary model, then a final blended gravitational
reference. The blended pair approximately preserves inner observed
rotation while its outer density approaches the measured \hihe profile.

Its substantially lower inferred outer density and gas-subtracted local
mass-to-light ratio illustrate the sensitivity of mass estimates to the assumed
gravitational contribution to gas rotation. This reduction follows from
our construction: without independent stellar kinematics, the blended
curve remains a model reference rather than a measured gravitational
rotation curve.

Assuming circular, axisymmetric midplane motion and a purely toroidal
field, we reconstructed the magnetic stress required by the observed--blended
rotation difference. The adopted isothermal pressure estimate gives only
a small correction over most of the outer region considered and was
therefore omitted from the field comparison.

The required field depends appreciably on vertical gas structure.
With hydrogen inertia, no pressure and $B_s=0$, constant exponential
heights from $1\kpc$ to $250\mathrm{pc}$ give peaks of approximately
$5.47$ to $10.95\mug$ near $34.9\kpc$; the additional $0.2\kpc$
case gives approximately $12.24\mug$. Extrapolated flaring exponential
and Gaussian profiles give approximately $5.25$ and $5.96\mug$,
respectively, near $33.8\kpc$. Including coupled helium raises these
amplitudes by approximately $15.5$ per cent. Field estimates must
therefore specify vertical structure, inertial density and boundary
condition.

Flaring solutions with $B_s=8\mug$ decline overall, with broad
intermediate variations, reaching approximately $5.73$ and $6.30\mug$
for exponential and Gaussian profiles, respectively, near the last
rotation measurement at $39.7\kpc$. The boundary term changes field
amplitude without changing the prescribed radial force residual.

The amplitudes and overall decline are qualitatively consistent with
the total field inferred by \citet{Basu2013}, decreasing outward to
approximately $8\mug$ near the mapped radio-disk edge. Non-zero-boundary
solutions provide possible outer continuations of this trend.
However, these radio estimates cover a different radial range and
determine neither the ordered azimuthal field nor its adopted boundary
value. The comparison supports the field scale but does not establish
a magnetic origin for the rotation-curve difference.

Separate stellar and gas kinematics, with constraints on outer field
strength, radial gradient and geometry, would test this interpretation.
The approach also applies to other galaxies with extended \hi disks,
quantifying the magnetic stresses required by specified gravitational
references.

\section*{Data availability}
The observational inputs are available in \citet{Battaglia2006} and
\citet{Kent1987}. 
%The digitised profiles, model data and analysis code
%are available from the corresponding author on reasonable request.

\bibliographystyle{mnras}
\bibliography{2026magnetic-biblio}
\bsp
\label{lastpage}
\end{document}